\documentclass{llncs}
\usepackage[T1]{fontenc} 
\usepackage{graphicx}
\usepackage[boxed,vlined,ruled,linesnumbered]{algorithm2e}
\usepackage{cite}
\usepackage{amsmath}
\usepackage{amsfonts}
 
\usepackage{color}
\usepackage{multicol}
\usepackage{multirow}
\usepackage{colortbl}
\usepackage{enumerate}

\begin{document}

\title{Optimising Problem Formulation for \\ Cylindrical Algebraic Decomposition
\thanks{The final publication is available at \texttt{http://link.springer.com}.}
}
\author{Russell Bradford \and James H. Davenport \and Matthew England \and David Wilson}
\institute{
University of Bath, Bath, BA2 7AY, U.K. \\
\email{ {\tt \{R.J.Bradford, J.H.Davenport, M.England, D.J.Wilson\}@bath.ac.uk}},\\ 
WWW home page: \texttt{http://people.bath.ac.uk/masjhd/Triangular/}
}
\maketitle

\begin{abstract}
Cylindrical algebraic decomposition (CAD) is an important tool for the study of real algebraic geometry with many applications both within mathematics and elsewhere.  It is known to have doubly exponential complexity in the number of variables in the worst case, but the actual computation time can vary greatly.  It is possible to offer different formulations for a given problem leading to great differences in tractability.  In this paper we suggest a new measure for CAD complexity which takes into account the real geometry of the problem.  This leads to new heuristics for choosing: the variable ordering for a CAD problem, a designated equational constraint, and formulations for truth-table invariant CADs (TTICADs).  We then consider the possibility of using Gr\"obner bases to precondition TTICAD and when such formulations constitute the creation of a new problem.


\end{abstract}


\section{Introduction} \label{SEC:Intro}

Cylindrical algebraic decomposition (CAD) is a key tool in real algebraic geometry both for its original motivation, quantifier elimination (QE) problems \cite[etc.]{Collins75}, but also in other applications ranging from robot motion planning \cite[etc.]{SS83II} to programming with complex functions \cite[etc.]{DBEW12} and branch cut analysis \cite[etc.]{EBDW13}. Decision methods for real closed fields are used in theorem proving \cite{DSW}, so CAD has much potential here.
 In particular  MetiTarski employs QEPCAD \cite{Brown03b} to decide statements in special functions using polynomial bounds  \cite{AP2008, AP2010, PPM}. Work is ongoing to implement a verified CAD procedure in {\sc Coq} \cite{M07,CM12}.

Since its inception there has been much research on CAD.  New types of CAD and new algorithms have been developed, offering improved performance and functionality.  The thesis of this paper is that more attention should now be given to how problems are presented to these algorithms. 


How a problem is formulated can be of fundamental importance to algorithms, rendering simple problems infeasible and vice versa.  In this paper we take some steps towards better formulation by introducing a new measure of CAD complexity and new heuristics for many of the choices required by CAD algorithms. We also further explore  preconditioning the input via Gr\"obner bases.  

\newpage

\subsection{Background on CAD}

A CAD is a decomposition of $\mathbb{R}^n$ into cells arranged cylindrically (meaning their projections are equal or disjoint) and described by semi-algebraic sets.  Traditionally CADs are produced sign-invariant to a given set of polynomials in $n$ variables ${\bf x}$, meaning the sign of the polynomials does not vary on the cells.  This definition was provided by Collins in \cite{Collins75} along with an algorithm which proceeded in two main phases.  The first, \textit{projection}, applies a projection operator repeatedly to a set of polynomials, each time producing another set of polynomials in one fewer variables.  Together these sets provide the {\em projection polynomials}.
The second phase, \textit{lifting}, then builds the CAD incrementally from these polynomials.  First $\mathbb{R}$ is decomposed into cells which are points and intervals corresponding to the real roots of the univariate polynomials.  Then $\mathbb{R}^2$ is decomposed by repeating the process over each cell using the bivariate polynomials at a sample point of the cell.  The output for each cell consists of {\em sections} of polynomials (where a polynomial vanishes) and {\em sectors} (the regions between these). Together these form the  {\em stack} over the cell, and taking the union of these stacks gives the CAD of $\mathbb{R}^2$.  This process is repeated until a CAD of $\mathbb{R}^n$ is produced. This final CAD will have cells ranging in dimension from 0 (single points) to $n$ (full dimensional portions of space).  The cells of dimension $d$ are referred to as {\em $d$-cells}.


It has often been noted that such decompositions actually do much more work than is required for most applications, motivating theory which considers not just polynomials but their origin.  For example, partial CAD \cite[etc.]{CH91} avoids unnecessary lifting over a cell if the solution to the QE problem on a cell is already apparent.  Another example is the use of CAD with equational constraints \cite[etc.]{McCallum99} where sign-invariance 
is only ensured over the sections of a designated equation, thus reducing the number of projection polynomials required.  It is worth noting that while the lifting stage takes far more resources that the projection, improvements of the projection operator have offered great benefits. 
   
Applications often analyse formulae (boolean combinations of polynomial equations, inequations and inequalities) by constructing a sign invariant CAD for the polynomials involved. However this analyses not only the given problem, but any formula built from these polynomials.  
In \cite{BDEMW13} the authors note that it would be preferable to build CADs directly from the formulae and so define a Truth Table Invariant CAD (TTICAD) as one which is has invariant truth values of various quantifier-free formulae (QFFs) in each cell. In \cite{BDEMW13} an algorithm was produced which efficiently constructed such objects for a wide class of problems by utilising the theory of equational constraints. 


\subsection{Formulating problems for CAD algorithms}

The TTICAD algorithm in \cite{BDEMW13} takes as input a sequence of QFFs, each of which is a formula with a designated equational constraint (an equation logically implied by the formula).  It outputs a CAD such that on each cell of the decomposition each QFF has constant truth value.  The algorithm is more efficient than constructing a full sign-invariant CAD for the polynomials in the QFFs, since it uses the theory of equational constraints for each QFF to reduce the projection polynomials used and hence the number of cells required.  Its benefit over equational constraints alone is that it may be used for formulae which do not have an overall explicit equational constraint (and to greater advantage than the use of implicit equational constraints). 
Many applications present problems in a suitable form for TTICAD, such as problems from branch cut analysis \cite{EBDW13}.

However, it is possible to envision problems where although separate QFFs are not imposed 
they could still lead to more economical CADs, (see Example \ref{ex:TTI-WE}).  
Further, 
we may consider splitting up individual QFFs if more than one equational constraint is present.  This leads to the question of how best to formulate the input to TTICAD, a question which motivated this paper and is answered in Section \ref{SEC:TTICAD}.  Some of this analysis could equally be applied to the theory of equational constraints alone and so this is considered in Section \ref{SEC:EqConst}. 

In devising heuristics to guide this process we realised that the existing measures for predicting CAD complexity could be misled.  An important use for these is choosing a variable ordering for a CAD; a choice which can make a substantial difference to the tractability of problems. We use $x \prec y$ to indicate $x$ is less than $y$ in an ordering.  In \cite{DSS04} the authors presented measures for CAD complexity but none of these consider aspects of the problem sensitive to the domain we work in (namely real geometry rather than complex).  In Section \ref{SEC:VarOrd} we suggest a simple new measure (the number of zero cells in the induced CAD of $\mathbb{R}^1$) leading to a new heuristic for use in conjunction with \cite{DSS04}.  We demonstrate in general it does well at discriminating between variable choices, and for certain problems is more accurate than existing heuristics.

These three topics are all examples of choices for the formulation of problems for CAD algorithms.  They are presented  in the opposite order to which they were considered above, as it is more natural for presenting the theory.  Problem formulation was considered in this conference series last year \cite{WBD12_GB} where the idea of preconditioning CAD using Gr\"obner bases was examined. This work is continued in Section \ref{SEC:Grobner} where we now consider preconditioning TTICAD.  

The tools developed for the formulation of input here lead to the question of whether their use is merely an addition to the algorithm or leads to the creation of a new problem.  This question also arose in \cite{WBD12_EX} where a project collecting together a repository of examples for CAD is described.  In Section \ref{SEC:Conclusion} we give our thoughts on this along with our conclusions and ideas for future work.  

\section{Choosing a Variable Ordering for CAD} \label{SEC:VarOrd}

\subsection{Effects of variable ordering on CAD}

It is well documented \cite[etc.]{DSS04} that the variable ordering used to construct a CAD can have a large impact on the number of cells and computation time. Example \ref{ex:WE1} gives a simple illustration.  Note that the effect of the variable ordering can be far greater than the numbers presented here and can change the feasibility of a given problem.  In \cite{BD07} the authors prove there are problems where one variable ordering will lead to a CAD with a constant number of cells while another will give a number of cells doubly exponential in the number of variables.

\begin{figure}[ht] 
\begin{center}
\includegraphics[width=4.5cm]{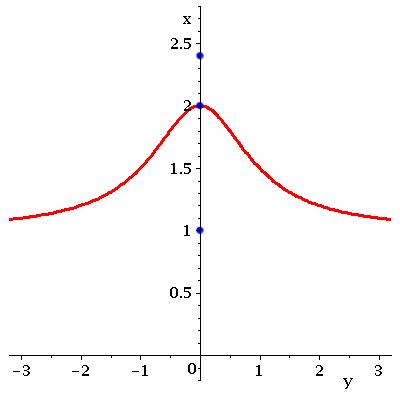}
\hspace*{0.3cm}
\includegraphics[width=4.5cm]{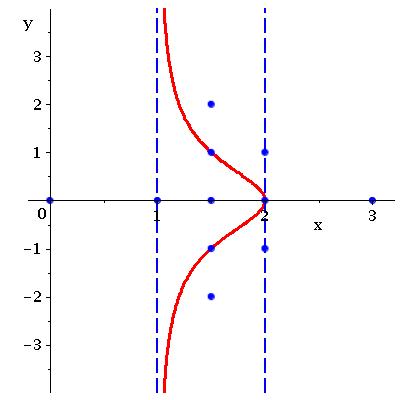}
\end{center}
\caption{Plots visualising the CADs described in Example \ref{ex:WE1}.  
} 
\label{fig:WE}
\end{figure}

\begin{example} \label{ex:WE1}
Consider the polynomial
$
f := (x-1)(y^2+1) - 1
$
whose graph is the solid curve in Figure \ref{fig:WE}.  We have two choices of variable ordering, which lead to the two different CADs visualised.  Each cell is indicated by a sample point (the solid circles).  Setting $y \prec x$ we obtain a CAD with 3 cells; the curve itself and the portions of the plane either side.  However, setting $x \prec y$ leads to a CAD with 11 cells; five 2-cells, five 1-cells and one 0-cell.  The dotted lines indicate the stacks over the 0-cells in the induced CAD of $\mathbb{R}^1$.  With $y \prec x$ the CAD of $\mathbb{R}^1$ had just one cell (the entire real line) while with $x \prec y$ there are five cells.

We note that these numbers occur using various CAD algorithms.  Indeed, for this simple example it is clear that these CADs are both minimal for their respective variable orderings, (i.e. there is no other decomposition which could have less cells whilst maintaining cylindricity.)
\end{example}

\subsection{Heuristics for choosing variable ordering}

In \cite{DSS04} the authors considered the problem of choosing a variable ordering for CAD and QE via CAD.  They identified a measure of CAD complexity that was correlated to the computation time, number of cells in the CAD and number of leaves in a partial CAD.  They identified the \emph{sum of total degrees of all monomials of all projection polynomials}, known as \texttt{sotd} and proposed the heuristic of picking the ordering with the lowest \texttt{sotd}.  Although the best known heuristic, \texttt{sotd} does not always pick the ideal ordering as demonstrated by some experiments in \cite{DSS04} and sometimes cannot distinguish between orderings as shown in Example \ref{ex:WE2}.

\begin{example} \label{ex:WE2}
Consider again the problem from Example \ref{ex:WE1}.  Applying any known valid projection operator to $f$ gives, with respect to $y$, the set of projection factors
$
\{ x-1, x-2 \},
$
(arising from the coefficients and discriminant of $f$).  Similarly, applying a projection operator with respect to $x$ gives
$
\{y^2+1\}.
$
Hence in this case both variable orderings have the same \texttt{sotd}.
\end{example}

We consider why \texttt{sotd} cannot differentiate between the orderings in this case. Algebraically, the only visible difference is that one ordering offers two factors of degree one while the other offers a single factor of degree two.
From Figure \ref{fig:WE} we see that one noticeable difference between the variable orderings is the number of 0-cells in the CAD of $\mathbb{R}^1$ (the dotted lines).  This is a feature of the real geometry of the problem as opposed to properties of the algebraic closure, measured by $\texttt{sotd}$.  
Investigating examples of this sort we devised a new measure \texttt{ndrr} defined to be the \emph{number of distinct real roots of the univariate projection polynomials} and created the associated heuristic of picking the variable ordering with lowest \texttt{ndrr}.  Considering again the projection factors from Example \ref{ex:WE2} we see that this new heuristic will correctly identify the ordering with the least cells.

The number of real roots can be identified, for example, using the theory of Sturm chains.  This extra calculation will likely take more computation time than the measuring of degrees required for \texttt{sotd}.  However, both costs are usually negligible compared to the cost of lifting in the CAD algorithm.

\subsection{Relative merits of the heuristics}

We do not propose \texttt{ndrr} as a replacement for \texttt{sotd} but suggest they are used together since both have relative merits.  We have already noted that the strength of \texttt{ndrr} is its ability to give information on the real geometry of the CAD.  Its weakness is that it only gives information on the complexity of the univariate polynomials, compared to \texttt{sotd} which measures at all levels.  If the key differences between orderings are not apparent in the univariate polynomials then \texttt{ndrr} is of little use, as in Example \ref{ex:quartic}.

\begin{example}
\label{ex:quartic}
Consider the problem of finding necessary and sufficient conditions on the coefficients of a quartic polynomial so that it is positive semidefinite:  eliminate the quantifier in,
$
\forall x (px^2+qx+r+x^2 \geq 0).
$  
This classic QE problem was first proposed in \cite{Lazard88} and was a test case in \cite{DSS04}.  There are six admissible variable orderings (since $x$ must always be projected first).  In all of these orderings the univariate projection factor set will consist of just the single variable of lowest order, (either $p,q$ or $r$) and hence all orderings will have an \texttt{ndrr} of one.  However, the \texttt{sotd} can distinguish between the orderings as reported in \cite{DSS04}.  
\end{example}

Despite the shortcoming of only considering the first level, \texttt{ndrr} should not be dismissed as effects at the bottom level can be magnified.  
We suggest using the heuristics in tandem, either using one to break ties between orderings which the other cannot discriminate or by taking a combination of the two measures. 

In \cite{DSS04} the authors suggested a second heuristic, a greedy algorithm based on \texttt{sotd}.  This approach avoided the need to calculate the projection polynomials for all orderings, instead choosing one variable at a time using the sum of total degree of the monomials from those projection polynomials obtained so far.  Unfortunately there is not an obvious greedy approach to using \texttt{ndrr}.  For problems involving many variables (so that calculating the full set of projection polynomials for each ordering is infeasible) we should revert to the \texttt{sotd} greedy algorithm, perhaps making use of \texttt{ndrr} to break ties.

\subsection{Coupled variables}

It has been noted in \cite{Phisanbut11} that a class of problems particularly unsuitable for \texttt{sotd} is choosing between coupled variables (two variables which are the real and imaginary parts of a complex variable). These are used, for example, when analysing complex functions by constructing a CAD to decompose the domain according to their branch cuts.  The ordering of the coupled variables for the CAD can affect the efficiency of the algorithm, as in Example \ref{ex:BC}.

\begin{figure}[ht] 
\begin{center}
\includegraphics[width=4.5cm]{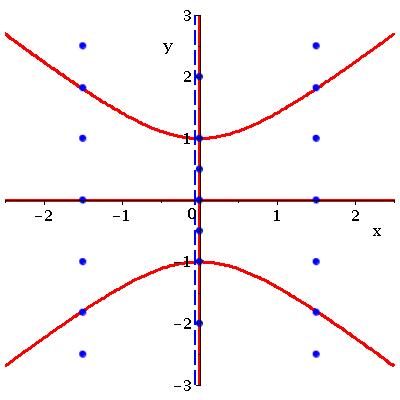}
\hspace*{0.3cm}
\includegraphics[width=4.5cm]{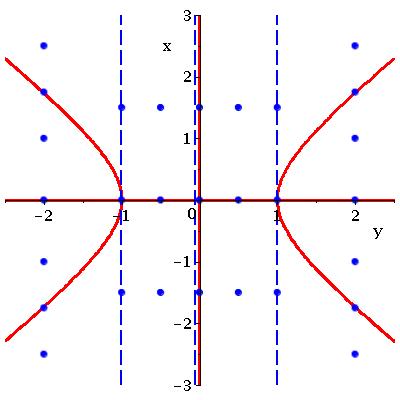}
\end{center}
\caption{Plots visualising the CADs described in Example \ref{ex:BC}.  
} 
\label{fig:Coupled}
\end{figure}

\begin{example}
\label{ex:BC}
Consider $f=\sqrt{z^2+1}$ where $z \in \mathbb{C}$.  The square root function has a branch cut along the negative real axis and so $f$ has branch cuts when
\begin{align*}
\Re(z^2+1) = x^2 - y^2 + 1 < 0 \quad \mbox{and} \quad \Im(z^2+1) = 2xy = 0,
\end{align*}
where $x,y$ are coupled real variables such that $z=x+{\rm{i}}y$.  With variable ordering $x \prec y$ we have ${\tt sotd} = 8, {\tt ndrr} = 4$ and a CAD with 21 cells while with variable ordering $y \prec x$ we have ${\tt sotd} = 8, {\tt ndrr} = 5$ and a CAD with 29 cells.  The CADs are visualised in Figure \ref{fig:Coupled} using the same techniques as described for Figure \ref{fig:WE}.  
\end{example}

\section{Designating Equational Constraints} \label{SEC:EqConst}

An \textbf{equational constraint} is an equation logically implied by a formula.  The theory of equational constraints is based on the observation that the formula will be false for any cell in the CAD where the equation is not satisfied.  Hence the polynomials forming any other constraints need only be sign invariant over the sections of the equational constraint.  The observation was first made in \cite{Collins98} with McCallum providing the first detailed approach in \cite{McCallum99}.  Given a problem with an equational constraint McCallum suggested a reduced projection operator, which will usually result in far fewer projection factors and a simpler CAD.  

This approach has been implemented in \textsc{Qepcad}, a command line interface for quantifier elimination through partial CAD \cite{Brown03b}.
It can also be induced in any implementation of TTICAD as discussed in Section \ref{SEC:TTICAD}.  The use of equational constraints can offer increased choice for problem formulation beyond that of picking a variable order.  If a problem has more than one equational constraint then one must be \emph{designated} for use in the algorithm.  We propose simple heuristics for making this choice based on \texttt{sotd} and \texttt{ndrr}.

Let $P$ be the McCallum projection operator which, informally, is applied to a set of polynomials to produce the coefficients, discriminant and cross resultants.  The full technical details are available in \cite{McCallum88} and a validated algorithm was given in \cite{McCallum98}.  Note that implementations usually make some trivial simplifications such as removal of constants, exclusion of polynomials that are identical to a previous entry (up to constant multiple), and only including those coefficients which are really necessary for the theory to hold. 

Next, for some equational constraint $f$ let $P_f$ be the reduced projection operator relative to $f$ described in \cite{McCallum99}.  Informally, this consists of the coefficients and discriminant of $f$ together with the resultant of $f$ taken with each of the other polynomials.  This is used for the first projection, reverting to $P$ for subsequent projections.
We can then apply the \texttt{sotd} and \texttt{ndrr} measures to the sets of projection polynomials as a measure of the complexity of the CADs that would be produced.  We denote these values by \texttt{S} and \texttt{N} respectively and our heuristics are then to choose the equational constraint that minimises these values.  

We ran experiments to test the effectiveness of these heuristics using problems from the CAD repository described in \cite{WBD12_EX}\footnote{Freely available at \url{http://opus.bath.ac.uk/29503}}.  We selected those problems with more than one equational constraint, for which at least one of the choices is tractable.  The experiments were run in \textsc{Maple} using the \texttt{ProjectionCAD} package \cite{England13} and the results are displayed in Table \ref{tab:ECTests} with the cell count, computation time and heuristic values given for each problem and choice of equational constraint.

\begin{table}
\begin{center}
\begin{tabular}{|lr|lccccr|lccccr|lccccr|}
\hline
\multirow{2}{*}{\textbf{\quad Problem}} 
&\,&\,& \multicolumn{4}{c}{EC Choice 1}  
&\,&\,& \multicolumn{4}{c}{EC Choice 2}  
&\,&\,& \multicolumn{4}{c}{EC Choice 3}    & \,  \\
&&& Cells & Time & \texttt{S} & \texttt{N} 
&&& Cells & Time & \texttt{S} & \texttt{N} 
&&& Cells & Time & \texttt{S} & \texttt{N} & \\
\hline
\, Intersection A		
&&& 657   & 5.6   & 61  & 7  		
&&& 463   & 5.1   & 64  & 8		
&&& \cellcolor[gray]{0.8} 269   & \cellcolor[gray]{0.8} 1.3   & \cellcolor[gray]{0.8} \textbf{42}  & \cellcolor[gray]{0.8} \textbf{4}  		
& \\
\, Intersection B		
&&& 711   & 6.3   & 66  & 6   	
&&& 471   & 5.4   & 71  & 6		
&&& \cellcolor[gray]{0.8} 303   & \cellcolor[gray]{0.8} 1.1   & \cellcolor[gray]{0.8} \textbf{40}  & \cellcolor[gray]{0.8} \textbf{5}  		
& \\
\, Random A			
&&& \cellcolor[gray]{0.8} 375  & \cellcolor[gray]{0.8} 2.7  & \cellcolor[gray]{0.8} 81  & \cellcolor[gray]{0.8} 9  		
&&& 435   & 3.6   & \textbf{73}  & \textbf{8}		
&&& 425   & 2.8   & 80  & \textbf{8}  		
& \\
\, Random B			
&&& 1295 & 21.4 & 140 & 13 		
&&& \cellcolor[gray]{0.8} 477   & \cellcolor[gray]{0.8} 3.8   & \cellcolor[gray]{0.8} \textbf{84}  & \cellcolor[gray]{0.8} \textbf{9}		
&&& 1437  & 23.9  & 158 & 14 		
& \\
\, Sphere-Catastrophe	
&&& 285  & 2.0  & 61  & 7   	
&&& \cellcolor[gray]{0.8} 169   & \cellcolor[gray]{0.8} 1.0   & \cellcolor[gray]{0.8} \textbf{59}  & \cellcolor[gray]{0.8} \textbf{5}		
&&&       &       &     &    		
& \\
\, Arnon84-2			
&&& 39   & 0.1  & 54  & 5  		
&&& \cellcolor[gray]{0.8} 9     & \cellcolor[gray]{0.8} 0.0   & \cellcolor[gray]{0.8} \textbf{47}  & \cellcolor[gray]{0.8} \textbf{1}     
&&&       &       &     &    		
& \\
\, Hong-90				
&&& F     & -     & \textbf{14}  & \textbf{0}  		
&&& F     & -     & \textbf{14}  & \textbf{0} 	
&&& \cellcolor[gray]{0.8} 27    & \cellcolor[gray]{0.8} 0.1   & \cellcolor[gray]{0.8} \textbf{14}  & \cellcolor[gray]{0.8} \textbf{0}   		
& \\
\, Cyclic-3			
&&& \cellcolor[gray]{0.8} 57   & \cellcolor[gray]{0.8} 0.3  & \cellcolor[gray]{0.8} \textbf{32}  & \cellcolor[gray]{0.8} \textbf{3}  		
&&& 117   & 0.7   & 35  & \textbf{3} 	
&&& 119   & 0.6   & 36  & 4 		
& \\
\hline
\end{tabular}
\end{center}
\caption{Comparing the choice of equational constraint for a selection of problems. The lowest cell count for each problem is highlighted and the minimal values of the heuristics emboldened.}
\label{tab:ECTests}
\end{table}

The full details on the problems can be found in the repository. The examples each contain two or three equational constraints and the numbering of the choices in the table refers to the order the equational constraints are listed in the repository.  The variable orderings used were those suggested in the repository.  The time taken to calculate \texttt{S} and \texttt{N} for each problem was always less than $0.05$ seconds and so insignificant to the overall timings.

For each problem the equational constraint choice resulting in the lowest cell count and timing has been highlighted and the minimal values of the heuristics emboldened.  We can see that for almost all cases both the heuristics point to the best choice.  However, there is an example (Random A) where both point to an incorrect choice.  The heuristic based on \texttt{sotd} is more sensitive (because it measures at all levels) and as a result is sometimes more effective.  For example, it picks the appropriate choice for the Cyclic-3 example while the other does not.     

Although the \texttt{sotd} heuristic is superior for all these examples it can be misled by examples where the real geometry differs, as in Example \ref{ex:EC}.

\begin{example}
\label{ex:EC}
Consider the polynomials 
\begin{align*}
f &:= y^5 - 2y^3x + yx^2 + y = y(y^2-(x+\rm{i}))(y^2-(x-\rm{i})) \\
g &:= y^5 - 2y^3x + yx^2 - y = y(y^2-(x+1))(y^2-(x-1))
\end{align*}
along with the formula $f=0 \wedge g=0$ and variable ordering $x \prec y$.  We could use either $f$ or $g$ as an equational constraint when constructing a CAD.  We have
\[
\mbox{discrim}(f) = 256(x^2+1)^3, \qquad 
\mbox{discrim}(g) = 256(x-1)^3(x+1)^3 
\]
and so both the projection sets have the same \texttt{sotd}.  However, with $f$ as an equational constraint the projection set has \texttt{ndrr}$=0$ while with $g$ it is $2$.  The CADs of $\mathbb{R}^2$ have 3 and 31 cells respectively.
\end{example}

\section{Formulating Input for TTICAD} \label{SEC:TTICAD}

Let $\Phi$ represent a set of QFFs, $\{\phi_i\}$. In \cite{BDEMW13} the authors define a Truth-Table Invariant CAD (TTICAD) as a CAD such that the boolean value of each $\phi_i$ is constant (either true or false) on each cell.  Clearly such a CAD is sufficient for solving many problems involving the formulae.  

A sign-invariant CAD is also a TTICAD, however, in \cite{BDEMW13} the authors present an algorithm to construct TTICADs more efficiently for the case where each $\phi_i$ has a designated equational constraint $f_i$ (an equation logically implied by $\phi_i$).  They adapt the theory of equational constraints to define a TTICAD projection operator and prove a key theorem explaining when it is valid.  Informally, the TTICAD projection operator produces the union of the application of the equational constraints projection operator to each $\phi_i$ along with the cross resultants of all the designated equational constraints, (see \cite{BDEMW13} for the full technical details).  As noted in the introduction, TTICAD is more efficient than equational constraints alone.

If there is more than one equational constraint present within a single $\phi_i$ then a choice must be made as to which is designated for use in the algorithm, (the others would then be treated as any other constraint).  As with choosing equational constraints in Section 3 the two different projection sets could be calculated and the measures $\texttt{sotd}$ and $\texttt{ndrr}$ taken and used as heuristics, picking the choice that leads to the lowest values.  

However, this situation actually offers further choice for problem formulation than the designation.  If $\phi_i$ had two equational constraints then it would be admissible to split this into two QFFs $\phi_{i,1}, \phi_{i,2}$ with one equational constraint assigned to each and the other constraints partitioned between them in any manner.  (Admissible because any TTICAD for $\phi_{i,1}, \phi_{i,2}$ is also a TTICAD for $\phi_i$.)  This is a generalisation of the following observation:  given a formula $\phi$ with two equational constraints a CAD could be constructed using either the traditional theory of equational constraints or the TTICAD algorithm applied to two QFFs.  On the surface it is not clear why the latter option would ever be chosen since it would certainly lead to more projection polynomials after the first projection.  However, a specific equational constraint may have a comparatively large number of intersections with another constraint, in which case, while separating these into different QFFs would likely increase the number of projection polynomials it may still reduce the number of cells in the CAD, (since the resultants taken would be less complicated leading to fewer projection factors at subsequent steps).  Example \ref{ex:TTI-WE} describes a simple problem which could be tackled using the theory of equational constraints alone, but for which it is beneficial to split into two QFFs and tackle with TTICAD.  

\begin{figure}[ht] 
\begin{center}
\includegraphics[width=4.5cm]{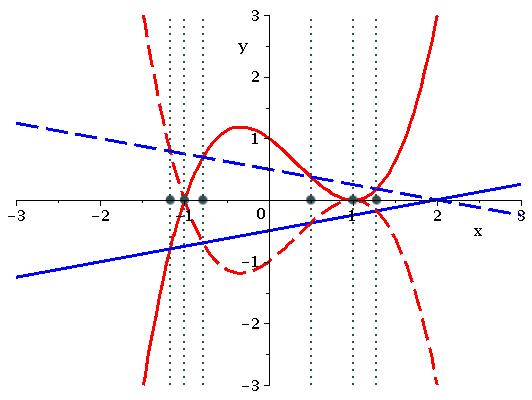}
\hspace*{0.3cm}
\includegraphics[width=4.5cm]{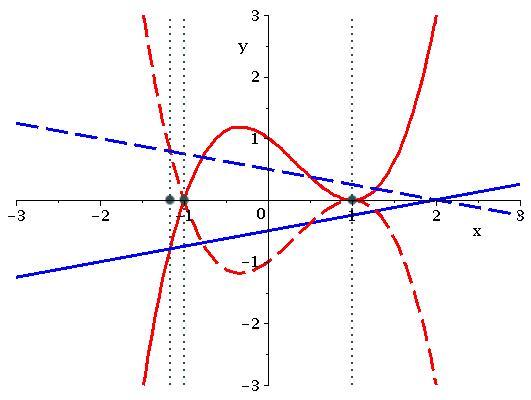}
\end{center}
\caption{Plots visualising the induced CADs of $\mathbb{R}^1$ described in Example \ref{ex:TTI-WE}.  } 
\label{fig:TTI-WE}
\end{figure}

\begin{example} \label{ex:TTI-WE}
Let $x \prec y$ and consider the polynomials
\begin{align*}
f_1 &:= (y-1) - x^3+x^2+x, \qquad \quad
g_1 := y - \tfrac{x}{4}+\tfrac{1}{2}, \\
f_2 &:= (-y-1) - x^3+x^2+x, \qquad \,
g_2 := -y - \tfrac{x}{4}+\tfrac{1}{2},
\end{align*} 
and the formula
$
\phi := f_1 = 0 \wedge g_1>0 \wedge f_2 = 0 \wedge g_2<0.
$

The polynomials are plotted in Figure \ref{fig:TTI-WE} where the solid curve is $f_1$, the solid line $g_1$, the dashed curve $f_2$ and the dashed line $g_2$.  The three figures also contain dotted lines indicating the stacks over the 0-cells of the CAD of $\mathbb{R}^1$ arising from the decomposition of the real line using various CAD algorithms.  

First, if we use the theory of equational constraints (with either $f_1$ or $f_2$ as the designated equational constraint) then a CAD is constructed which identifies all the roots and intersection between the four polynomials except for the intersection of $g_1$ and $g_2$.  (Note that this would be identified by a full sign-invariant CAD).  This is visualised by the plot on the left while the plot on the right relates to a TTICAD with two QFFs.  In this case only three 0-cells are identified, with the intersections of $g_2$ with $f_1$ and $g_1$ with $f_2$ ignored.

The TTICAD has 31 cells while the CADs produced using equational constraints both have 39 cells.  The TTICAD projection set has an \texttt{sotd} of 26 and an \texttt{ndrr} of $3$ while each of the CADs produced using equational constraints have projection sets with values of 30 and 6 for \texttt{sotd} and \texttt{ndrr}. 
\end{example}


As suggested by Example \ref{ex:TTI-WE} we propose using the measures \texttt{sotd} and \texttt{ndrr} applied to the set of projection polynomials as heuristics for picking an approach.  We can apply these with the TTICAD projection operator for deciding if it would be beneficial to split QFFs.  This can also be used for choosing whether to use TTICAD instead of equational constraints alone, since applying the TTICAD algorithm from \cite{BDEMW13} on a single QFF is equivalent to creating a CAD invariant with respect to an equational constraint.

We may also consider whether it is possible to combine any QFFs.  If the formulae were joined by conjunction then it would be permitted and probably beneficial but we would then need to choose which equational constraint to designate.  Formulae joined by disjunction could also be combined if they share an equational constraint, (with that becoming the designated choice in the combined formula).  Such a situation is common for the application to branch  cut analysis since many branch cuts come in pairs which lie on different portions of the same curve.  However, upon inspection of the projection operators, we see that such a merger would not change the set of projection factors in the case where the shared equational constraint is the designated one for each formula.  Note, if the shared equational constraint is not designated in both then the only way to merge would be by changing designation. 

When considering whether to split and which equational constraint to designate the number of possible formulations increases quickly.  Hence we propose a method for TTICAD QFF formulation, making the choices one QFF at a time.  Given a list $\hat{\Phi}$ of QFFs (quantifier free formulae):
\begin{enumerate}[(1)]
\item Take the disjunction of the QFFs and put that formula into disjunctive normal form, $\bigvee \hat{\phi}_i$ so that each $\hat{\phi}_i$ is a conjunction of atomic formulae.
\item Consider each $\hat{\phi}_i$ in turn and let $m_i$ be the number of equational constraints.  
\begin{itemize}
\item If $m_i=0$ then $\hat{\Phi}$ is not suitable for the TTICAD algorithm of \cite{BDEMW13}, (although we anticipate that it could be adapted to include such cases).  
\item If $m_i=1$ then the sole equational constraint is designated trivially.  
\item If $m_i>1$ then we consider all the possible partitions of the formula in $\hat{\phi}_i$ into sub QFFs with at least one equational constraint each, and all the different designations of equational constraint within those sub-QFFs with more than one.  Choose a partition and designation for this clause according to the heuristics based on \texttt{sotd} and \texttt{ndrr} applied to the projections polynomials from the clause.
\end{itemize}
\item Let $\Phi$ be the list of new QFFs, $\phi_i$, and the input to TTICAD.
\end{enumerate}

\section{Using Gr\"obner Bases to Precondition TTICAD QFFs} \label{SEC:Grobner}

Recall that for an ideal, $I \subset \mathbb{R}[{\bf x}]$, a {\em Gr\"obner basis} (for a given monomial ordering) is a  polynomial basis of $I$ such that $\{ {\rm lm}(g) \mid g \in G\}$ is also a basis for $\{ {\rm lm}(f) \mid f \in I\}$. 
In \cite{BH91} experiments were conducted to see if Gr\"obner basis techniques could precondition problems effectively for CAD. Given a problem:
\begin{equation*}\textstyle
\varphi:= \bigwedge_{i=1}^s f_i({\bf x}) = 0,
\end{equation*}
a purely lexicographical Gr\"obner basis $\{\hat{f}_i\}_{i=1}^t$ for the $f_i$, (taken with respect to the same variable ordering as the CAD), could take their place to form an equivalent sentence:
\begin{equation*}\textstyle
\hat{\varphi}:= \bigwedge_{i=1}^t \hat{f}_i({\bf x}) = 0.
\end{equation*}
Initial results suggested that this preconditioning can be hugely beneficial in certain cases, but may be disadvantageous in others.

In \cite{WBD12_GB} this idea was considered in greater depth. A larger base of problems was tested and the idea extended to include Gr\"obner reduction. Given a problem: 
\begin{equation*}\textstyle
\psi:= \left( \bigwedge_{i=1}^{s_1} f_i({\bf x}) = 0 \right) \land \left( \bigwedge_{i=1}^{s_2} g_i({\bf x}) \ast_i 0 \right), \qquad \ast_i \in \{=,\neq,>,<\},
\end{equation*}
 you can first compute $\{ \hat{f}_i\}_{i=1}^{t_1}$ followed by reducing the $g_i$ with respect to the $\hat{f}_i$ to obtain $\{ \hat{g}_i\}_{i=1}^{t_2}$. Then the following sentence will be equivalent to $\psi$:
\begin{equation*}\textstyle
\hat{\psi}:= ( \bigwedge_{i=1}^{t_1} \hat{f}_i({\bf x}) = 0 ) \land ( \bigwedge_{i=1}^{t_2} \hat{g}_i({\bf x}) \ast_i 0 ).
\end{equation*}

Experimentation showed that this Gr\"obner preconditioning can be highly beneficial with respect to both computation time and cell count, however the effect is not universal. To identify when preconditioning is beneficial a simple metric was posited and shown to be a good indicator. The quantity {\tt TNoI} ({\em total number of indeterminates}) for a set of polynomials $F$ is simply defined to be the sum of the number of variables present in each polynomial in $F$. In all testing carried out (both for \cite{WBD12_GB} and henceforth) if the produced Gr\"obner basis has a lower {\tt TNoI} than the original set of polynomials then preconditioning is beneficial for sign-invariant CAD (the converse is not always true). 

A natural question is whether Gr\"obner preconditioning can be adapted for TTICAD.  This is possible by performing the Gr\"obner preconditioning on the individual QFFs. There is a necessity, however, for a problem to be suitably complicated for this preconditioning to work: each QFF must have multiple equational constraints amenable to the creation of a Gr\"obner Basis. 
This required complexity means there are few examples in the literature which are suitable and tractable for experimentation.  We demonstrate the power of combining these two techniques through a worked example.

\begin{figure}[t] 
\begin{center}
\includegraphics[width=5cm]{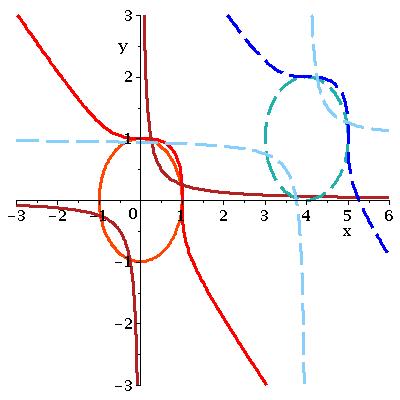}
\end{center}
\caption{Plot of the functions described in Example \ref{ex:TTI-Gro}.} 
\label{fig:TTI-Gro1}
\end{figure}

\begin{example}\label{ex:TTI-Gro}
Consider the polynomials
\begin{align*}
\begin{array}{rclcrcl}
f_{1,1} &:=& {x}^{2}+{y}^{2}-1, & \qquad \qquad & f_{2,1} &:=& \left( x-4 \right) ^{2} + \left( y-1 \right) ^{2} - 1, \\
f_{1,2} &:=& {x}^{3}+{y}^{3}-1, & &
f_{2,2} &:=& \left( x-4 \right) ^{3} + \left( y-1 \right) ^{3} - 1, \\
g_1 &:=& xy - \tfrac{1}{4}, & & g_2 &:=& \left( x-4 \right)  \left( y-1 \right) - \tfrac{1}{4}
\end{array}
\end{align*} 
and the formula $\left[ f_{1,1} = 0 \land f_{1,2} = 0  \land g_1 > 0 \right] \lor \left[ f_{2,1} = 0 \land f_{2,2} = 0  \land g_2 > 0 \right]$.

The polynomials are plotted in Figure \ref{fig:TTI-Gro1} where the solid curves represent $f_{1,1},f_{1,2},g_1$, and the dashed curves $f_{2,1},f_{2,2},g_2$.
\end{example}

%

We will consider both variable orderings: $y \prec x$ and $x \prec y$. We can compute full CADs for this problem, with 725 and 657 cells for the respective orderings.  If we use TTICAD to tackle the problem then there are four possible two-QFF formulations, (splitting QFFs is not beneficial for this problem). The four formulations are described in the second column of Table \ref{tab:TTI-Gro}.

We can apply Gr\"obner preconditioning to both QFFs separately, computing a Gr\"obner basis, with respect to the compatible ordering, of $\{ f_{i,1},f_{i,2}\}$. For both QFFs and both variable orderings three polynomials are produced. We denote them by  $\{ \hat{f}_{i,1},\hat{f}_{i,2}, \hat{f}_{i,3}\}$ (note the polynomials differ depending on the variable ordering). The algorithm used to compute these bases gives the polynomials in decreasing order of leading monomials with respect to the order used to compute the basis (purely lexicographical).

\begin{table}[t]
\begin{center}
\begin{tabular}{|c|cc||c|cc|cc||c|cc|cc|}
\hline
\multirow{2}{*}{Order} 
       & \multicolumn{2}{c||}{Full CAD} &  \multicolumn{5}{c||}{TTI CAD} & \multicolumn{5}{c|}{TTI+Gr\"{o} CAD} \\
       & Cells & Time & Eq Const & Cells & Time & {\tt S} & {\tt N} & Eq Const & Cells & Time & {\tt S} & {\tt N}\\\hline
 $y \prec x$ 
       & 725 & 22.802 & $f_{1,1},f_{2,1}$ & 153 & 0.818 & {\bf 62} & 12 & \cellcolor[gray]{0.8} $\hat{f}_{1,1},\hat{f}_{2,1}$ &  \cellcolor[gray]{0.8} 27 & \cellcolor[gray]{0.8} 0.095 &\cellcolor[gray]{0.8}  {\bf 37} & \cellcolor[gray]{0.8} {\bf 3} \\
       &     &        & $f_{1,1},f_{2,2}$ & 111 & 0.752 & 94 & 10 & $\hat{f}_{1,1},\hat{f}_{2,2}$ &  47 & 0.361 & 50 & 5 \\
       &     &        & $f_{1,2},f_{2,1}$ & 121 & 0.732 & 85 &  9 & $\hat{f}_{1,1},\hat{f}_{2,3}$ &  93 & 0.257 & 50 & 9 \\
       &     &        & \cellcolor[gray]{0.8} $f_{1,2},f_{2,2}$ & \cellcolor[gray]{0.8}  75 & \cellcolor[gray]{0.8} 0.840 & \cellcolor[gray]{0.8} 99 & \cellcolor[gray]{0.8}  {\bf 7} & $\hat{f}_{1,2},\hat{f}_{2,1}$ &  47 & 0.151 & 47 & 5 \\
       &     &        &                   &     &       &    &    & $\hat{f}_{1,2},\hat{f}_{2,2}$ &  83 & 0.329 & 63 & 7 \\ 
       &     &        &                   &     &       &    &    & $\hat{f}_{1,2},\hat{f}_{2,3}$ & 145 & 0.768 & 81 & 11 \\ 
       &     &        &                   &     &       &    &    & $\hat{f}_{1,3},\hat{f}_{2,1}$ &  95 & 0.263 & 46 & 10 \\ 
       &     &        &                   &     &       &    &    & $\hat{f}_{1,3},\hat{f}_{2,2}$ & 151 & 0.712 & 80 & 12 \\ 
       &     &        &                   &     &       &    &    & $\hat{f}_{1,3},\hat{f}_{2,3}$ & 209 & 0.980 & 62 & 16 \\ \hline
 
 $x \prec y$ 
       & 657 & 22.029 & $f_{1,1},f_{2,1}$ & 125 & 0.676 & {\bf 65}  & 14 & \cellcolor[gray]{0.8} $\hat{f}_{1,1},\hat{f}_{2,1}$ & \cellcolor[gray]{0.8}  29 & \cellcolor[gray]{0.8} 0.085 & \cellcolor[gray]{0.8} {\bf 39} & \cellcolor[gray]{0.8} {\bf 4} \\
       &     &        & $f_{1,1},f_{2,2}$ & 117 & 0.792 & 96  & 11 & $\hat{f}_{1,1},\hat{f}_{2,2}$ &  53 & 0.144 & 52 & 6 \\
       &     &        & $f_{1,2},f_{2,1}$ & 117 & 0.728 & 88  & 11 & $\hat{f}_{1,1},\hat{f}_{2,3}$ &  97 & 0.307 & 53 & 97 \\
       &     &        &\cellcolor[gray]{0.8}  $f_{1,2},f_{2,2}$ & \cellcolor[gray]{0.8}  85 & \cellcolor[gray]{0.8} 0.650 & \cellcolor[gray]{0.8} 101 & \cellcolor[gray]{0.8}  {\bf 8} & $\hat{f}_{1,2},\hat{f}_{2,1}$ &  53 & 0.146 & 49 & 6 \\
       &     &        &                   &     &       &     &    & $\hat{f}_{1,2},\hat{f}_{2,2}$ &  93 & 0.332 & 65 & 8 \\ 
       &     &        &                   &     &       &     &    & $\hat{f}_{1,2},\hat{f}_{2,3}$ & 149 & 0.782 & 81 & 13 \\
       &     &        &                   &     &       &     &    & $\hat{f}_{1,3},\hat{f}_{2,1}$ &  97 & 0.248 & 48 & 11 \\ 
       &     &        &                   &     &       &     &    & $\hat{f}_{1,3},\hat{f}_{2,2}$ & 149 & 0.798 & 82 & 13 \\  
       &     &        &                   &     &       &     &    & $\hat{f}_{1,3},\hat{f}_{2,3}$ & 165 & 1.061 & 65 & 18       
\\\hline
\end{tabular}
\end{center}
\caption{Experimental results relating to Example \ref{ex:TTI-Gro}. The lowest cell counts are highlighted and the minimal values of the heuristics emboldened.}\label{tab:TTI-Gro}
\end{table}

Table \ref{tab:TTI-Gro} shows that the addition of Gr\"obner techniques to TTICAD can produce significant reductions: a drop from 153 cells in 0.8s to 27 cells in under 0.1s (including the time required to compute the Gr\"obner bases). As discussed in \cite{WBD12_GB}, preconditioning is not always beneficial, as evident from the handful of cases that produce  more cells than TTICAD alone. As with Table \ref{tab:ECTests} we have highlighted the examples with lowest cell count and emboldened the lowest heuristic.
Looking at the values of {\tt S} and {\tt N} we see that for this example {\tt ndrr} is the best measure to use.  

In \cite{WBD12_GB} {\tt TNoI} was used to predict whether preconditioning by Gr\"obner Basis would be beneficial. In this example {\tt TNoI} is increased in both orderings by taking a basis, which correctly predicts a bigger full CAD after preconditioning. However, {\tt TNoI} does not take into account the added subtlety of TTICAD (as shown by the huge benefit above).


\section{Conclusions and Future Work} \label{SEC:Conclusion}

In this paper we have considered various issues based around the formulation of input for CAD algorithms.  We have revisited the classic question of choosing the variable ordering, proposing a new measure of CAD complexity \texttt{ndrr} to complement the existing \texttt{sotd} measure.  We then used these measures as heuristics for the problem of designating equational constraints and QFF formulation for TTICAD.  Finally we considered the effect of preconditioning by Gr\"obner bases.  

It is important to note that these are just heuristics and, as such, can be misleading for certain examples. Although the experimental results in Section \ref{SEC:EqConst} suggest {\tt sotd} is a finer heuristic than {\tt ndrr} we have demonstrated that there are examples when {\tt ndrr} performs better, not just Example \ref{ex:EC} which was contrived for the purpose but also Example \ref{ex:TTI-Gro} introduced for the work on Gr\"obner bases.

These issues have been treated individually but of course they intersect.  For example it is also necessary to pick a variable ordering for TTICAD.  This choice will need to made before employing the method for choosing QFF formulation described in Section \ref{SEC:TTICAD}.  However, the optimal choice of variable ordering for one QFF formulation may not be optimal for another!  For example, the TTICAD formulation with two QFFs was the best choice in Example \ref{ex:TTI-WE} where the variable ordering was stated as $x \prec y$ but if we had $y \prec x$ then a single QFF is superior.  

The idea of combining TTICAD with Gr\"obner preconditioning (discussed in \cite{BH91}, \cite{WBD12_GB}) is shown, by a worked example, to have the potential of being a very strong tool. However, this adds even more complication in choosing a formulation for the problem. 
Taken together, all these choices of formulation can become combinatorially overwhelming and so methods to reduce this, such as the greedy algorithm in \cite{DSS04} or the method at the end of Section \ref{SEC:TTICAD}, are of importance.

All these options for problem formulation motivate the somewhat philosophical question of when a reformulation results in a new problem.  When a variable ordering is imposed by an application (such as projecting quantified variables first when using CAD for quantifier elimination) then violating this would clearly lead to a new problem while changing the ordering within quantifier blocks could be seen to be a optimisation of the algorithm.  Similar distinctions could be drawn for other issues of formulation.

Given the significant gains available from problem reformulation it would seem that the existing technology could benefit from a redesign to maximise the possibility of its use. For example, CAD algorithms could allow the user to input the variables is quantifier blocks so that the technology can choose the most appropriate ordering that still solves the problem.

We finish with some ideas for future work on these topics.
\begin{itemize}
\item All the work in this paper has been stated with reference to CAD algorithms based on projection and lifting.  A quite different approach, CAD via Triangular Decomposition, has been developed in \cite{CMXY09} and implemented as part of the core \textsc{Maple} distribution.  This constructs a (sometimes quite different) sign-invariant CAD by transferring the problem to complex space for solving.  A key question is how much of the work here transfers to this approach?
\item Can the heuristics for choosing equational constraints also be used for choosing pivots when using the theory of bi-equational constraints in \cite{BM05}?
\item Can the \texttt{ndrr} measure be adapted to consider also the real roots of those projection polynomials with more than one variable?
\end{itemize}

We finish by discussing one of the initial motivations for engaging in work on problem formulation: a quantifier elimination problem proving a property of Joukowski's transformation. This is the transformation $z \mapsto \frac{1}{2}(z + \frac{1}{z})$ which is used in aerodynamics to create an aerofoil from the unit circle. The fact it is bijective on the upper half plane is relatively simple to prove analytically but we found the state of the art CAD technology was incapable of producing an answer in reasonable time.  Then, in a personal communication, Chris Brown described how reformulating the problem with a succession of simple logical steps makes it amenable to {\sc Qepcad}, allowing for a solution in a matter of seconds. These steps included splitting a disjunction to form two separate problems and the (counter-intuitive) removal of quantifiers which block {\sc Qepcad}'s use of equational constraints.  Further details are given in \cite[Sec. III]{DBEW12} and in the future we aim to extend our work on problem formulation to develop techniques to automatically render this problem feasible.

\subsection*{Acknowledgements}

This work was supported by the EPSRC grant: EP/J003247/1.  The authors would like to thank Scott McCallum for many useful conversations on TTICAD and Chris Brown for sharing his work on the Joukowski transformation.

\end{document}